\documentclass[journal,transmag]{IEEEtran}

\usepackage{cite}
\usepackage{amsmath,amssymb,amsfonts}
\usepackage{algorithmic}
\usepackage{graphicx}
\usepackage{textcomp}
\usepackage{xcolor}
\usepackage{todonotes}
\usepackage{bm}
\usepackage{xspace}
\usepackage{booktabs}
\usepackage{tikzscale}
\usepackage{tikz}
\usepackage{pgfplots}
\usepackage{inputenc}
\usepackage{standalone}
\usepackage{multirow}
\usepackage{pdfpages}
\usepackage[section]{placeins}
\usepackage{bm}
\usepackage{setspace}
\usepackage{soul}
\usepackage{cancel}
\usepackage{mathtools}

\newcommand{\threeD}{\mbox{3-D}\xspace}
\newcommand{\twoD}{\mbox{2-D}\xspace}
\newcommand{\hphi}{\mbox{H-$\phi$}\xspace}

\newlength\figH
\newlength\figW
\usepackage[colorlinks, linkcolor={red!50!black},citecolor={blue!50!black},urlcolor={blue!80!black}]{hyperref}

\begin{document}
	%
	\title{COMSOL implementation of the H-$\phi$-formulation with thin cuts for modeling superconductors with transport currents}

	
	\author{\IEEEauthorblockN{Alexandre Arsenault, Bruno de Sousa Alves, and Fr\'ed\'eric Sirois}
		\IEEEauthorblockA{ Polytechnique Montr\'eal, Montr\'eal, Canada}
		\thanks{Corresponding author: Alexandre Arsenault (alexandre-1.arsenault@polymtl.ca)}
	}
	
	\IEEEtitleabstractindextext{%
		\begin{abstract}
			Despite the acclaimed success of the magnetic field (H) formulation for modeling the electromagnetic behavior of superconductors with the finite element method, the use of vector-dependent variables in non-conducting domains leads to unnecessarily long computation times. In order to solve this issue, we have recently shown how to use a magnetic scalar potential together with the H-formulation in the COMSOL Multiphysics environment to efficiently and accurately solve for the magnetic field surrounding superconducting domains. However, from the definition of the magnetic scalar potential, the non-conducting domains must be made simply connected in order to obey Ampere's law. In this work, we use thin cuts to apply a discontinuity in $\phi$ and make the non-conducting domains simply connected. This approach is shown to be easily implementable in the COMSOL Multiphysics finite element program, already widely used by the applied superconductivity community. We simulate three different models in \twoD and \threeD using superconducting filaments and tapes, and show that the results are in very good agreement with the H-A and H-formulations. Finally, we compare the computation times between the formulations, showing that the \mbox{\hphi-formulation} can be up to seven times faster than the standard H-formulation in certain applications of interest.
		\end{abstract}
		
		}

	\maketitle

	\IEEEdisplaynontitleabstractindextext

	%
	\IEEEpeerreviewmaketitle

	\section{Introduction}
	%
	%
	%
	%
	\IEEEPARstart{A}{s} the development of novel high temperature superconductor (HTS) applications increases in complexity, so do the numerical models required to accurately predict their electromagnetic (EM) behavior. In turn, this complexity calls for more computational power and efficient methods in order to complete simulations in a reasonable amount of time. 
	
	One of the most frequently used methods to model the EM behavior of HTSs is the finite element method (FEM) solving the H-formulation \cite{shen2020,shen2020a}. This formulation combines Faraday and Ampere's laws to directly solve for the magnetic field as the dependent variable. Although it has proven to accurately model applications ranging from AC losses \cite{zhao2017,ainslie2012,grilli2014,brambilla2007} in HTS tapes to the magnetic levitation of permanent magnets over HTS bulks \cite{bernstein2020,zheng2021,sass2015,grilli2018}, the use of H in non-conducting domains is problematic. First, the dummy resistivity used to avoid eddy currents in non-conducting domains degrades the matrix conditioning \cite{arsenault2021} and leads to unwanted currents \cite{stenvall2014}. In addition, the use of a vector-dependent variable in these domains unnecessarily adds degrees of freedom (DOFs) to the model \cite{lahtinen2015}.
	
	The magnetic scalar potential $\phi$ is perfectly suited to solve the aforementioned issues when simulating HTSs. Solely using $\phi$ in non-conducting domains, the current density is automatically zero by definition. In addition, $\phi$ is a scalar, leading to a decrease in the number of DOFs and therefore in computation times.
	
	However, a major difficulty arising from the use of the magnetic scalar potential is that the domains simulated with $\phi$ must be simply connected in order to avoid violating Ampere's law, as described in the next section. This issue did not emerge in our recent works \cite{arsenault2021,arsenault2021a}, since the non-conducting domains did not surround any net current in the models considered. Nonetheless, in many applications of interest, a net current is needed in the HTS domains \cite{paul1997,zermeno2013,zermeno2013a,shajii1994,brouwer2019}. 
	
	Several methods have been proposed to make multiply connected domains simply connected. For example, in \cite{rodger1987}, two different methods are compared: adding a thin conducting layer in the air domains and imposing a discontinuity in $\phi$ along a ``thin cut" equal to the current inside the conductor. More recently, the use of edge-based cohomology basis functions or ``thick cuts" was shown to efficiently make multiply connected domains simply connected \cite{lahtinen2015,stenvall2014,gross1995,pellikka2013,dlotko2019}. Although this last method is very promising, a home-brewed or open-source FEM software must be programmed or utilized in order to generate the thick cuts. In addition, a good understanding of algebraic topology is required to fully comprehend the modeling details behind thick cuts.
	
	In this work, we impose a transport current in the \mbox{\hphi-formulation} by using thin cuts in the commercial FEM software COMSOL Multiphysics. We simulate three different models comprised of HTS filaments and tapes and evaluate the accuracy of the \mbox{\hphi-formulation} by comparing the same models computed with the standard H-formulation and with the H-A-formulation. All simulations are implemented in COMSOL Multiphysics 5.5 on a personal computer with an Intel(R) Core(TM) i7-3770 processor and 32~GB of random access memory.
	
	\section{Theory and model}
	
	The \mbox{\hphi-formulation} combines the H-formulation in conducting domains with a magnetic scalar potential formulation, $\phi$, in non-conducting domains. In the H-formulation, we solve Faraday's law in the form:
	\begin{equation}
	\quad \nabla\times \left(\rho\nabla\times\mathbf{H}\right) =-\mu_0\frac{\partial \mathbf{H}}{\partial t} \,,
	\label{eq:H}
	\end{equation}
	where $\rho$ is the resistivity and $\mu_0$ is the magnetic permeability in vacuum. The non-linear resistivity of HTSs is modeled using the power law model \cite{rhyner1993}:
	\begin{equation}
	\rho=\frac{E_c}{J_c}\left(\frac{|\mathbf{J}|}{J_c}\right)^{n-1},
	\label{eq:rho}
	\end{equation}
	where $\mathbf{J}$ is the current density, $J_c$ is the critical current density, $n$ is the critical exponent, and $E_c$ is the electric field criterion. In this work, $n=40$ and $E_c=1~\mu$V/cm are used for all the simulations.
	
	In non-conducting domains, the current density is equal to zero so that Ampere's law gives $\nabla\times\mathbf{H}=0$. Then, Gauss' law, $\nabla\cdot\mathbf{B}=0$, together with $\mathbf{B}=\mu_0\mathbf{H}$ and $\mathbf{H}=-\nabla\phi$ are combined to obtain the governing equation in non-conducting domains, i.e.
	\begin{equation}
	\nabla\cdot\nabla\phi=0.
	\label{eq:phi}
	\end{equation}
	Since no external field is applied, the magnetic flux through exterior boundaries is set to zero in order to completely define the problem.
	
	The coupling between the two physics is explained in detail in \cite{arsenault2021}. In short, the tangential component of the fields at the interface between H and $\phi$ are equated in the H physics, while the perpendicular components are equated in the $\phi$ physics. Through the couplings in both directions, the full vector field is defined at the interface between the two physics.

	The difficulty in modeling transport currents in the \mbox{\hphi-formulation} arises from the fact that Ampere's law is not naturally obeyed when a net current passes through a multiply-connected, non-conducting domain. This causes the magnetic scalar potential to be multi-valued, such that it cannot be uniquely solved. Several articles and books rigorously describe this phenomenon \cite{nicolet1995,meunier2008,verite1987,ren2002,kotiuga1987,leonard1989}, so we will only briefly explain it in this work. 
	
	First, recall the integral form of Ampere's law:
	\begin{equation}
	\oint_C\mathbf{H}\cdot \textrm{d}\mathbf{l}=I_{\textnormal{enc}},
	\label{eq:ampere}
	\end{equation}
	where the displacement currents are neglected and $I_{\textnormal{enc}}$ is the enclosed current inside loop $C$. In non-conducting domains, the current is zero by definition, so that any line integral over $C$ should be zero. However, this condition is not always satisfied, as illustrated below.
	
\begin{figure}[tb]
		\centering
		\includegraphics[width=\linewidth]{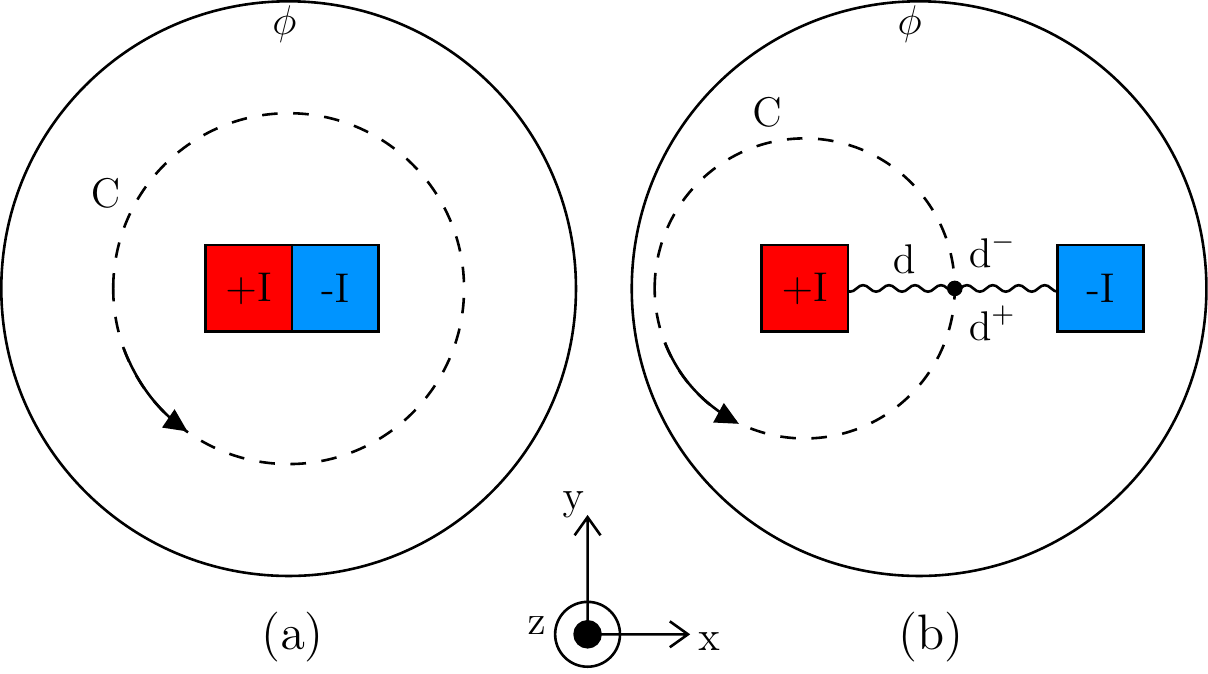}
		\caption{Cross section of two infinitely long rectangular conductors carrying current $I$ in opposite directions. (a) Current distribution obeying Ampere's law on any loop $C$ in the $\phi$ domain. (b) Current distribution violating Ampere's law on loop $C$ unless a discontinuity in $\phi$ is imposed on $d$. $d^-$ and $d^+$ represent points above and below the discontinuity, respectively.  Illustration created on \url{www.mathcha.io}.}
		\label{fig:ampere}
	\end{figure}

	Consider two infinitely long conductors carrying equal and opposite currents, as demonstrated by the \twoD cross section in Fig.~\ref{fig:ampere}. If the conductors are in contact with each other, as in Fig.~\ref{fig:ampere}(a), any loop $C$ in the $\phi$ physics will enclose zero net current. Thus, Ampere's law is naturally obeyed in this configuration. This explains why the \mbox{\hphi-formulation} can naturally be used to magnetize a superconducting bulk in a uniform field as in \cite{arsenault2021}, since the eddy currents yield a net current of zero inside the bulk. 
	
	Conversely, when the conductors are separated from each other as in Fig.~\ref{fig:ampere}(b), there exists a loop $C$ for which the enclosed current is non-zero, therefore contradicting the definition of $\phi$. 
	
	In order to solve this problem, we introduce a thin cut in the non-conducting domain to make it simply connected \cite{rodger1987}. In other words, we impose a discontinuity of $\phi$ on a line such that any loop $C$ in the $\phi$ physics obeys Ampere's law. By adding the discontinuity of $\phi$ on $d$ in Fig.~\ref{fig:ampere}(b), any loop integral of the magnetic field passing between the current carrying conductors will acquire a jump equal to the enclosed current. This can be mathematically described by using the gradient theorem, such that Ampere's law becomes:
	\begin{align}
	\oint_C\mathbf{H}\cdot \textrm{d}\mathbf{l}=-\oint_C\mathbf{\nabla}\phi\cdot \textrm{d}\mathbf{l}&=\phi(d^-)-\phi(d^+)\equiv [\phi]_d \coloneqq I_{\text{enc}},
	\label{eq:discontinuity}
	\end{align}
	where $\phi(d^+)$ and $\phi(d^-)$ are the magnetic scalar potentials evaluated on each side of the discontinuity. The notation $[\phi]_d$ is commonly used in the literature to represent the discontinuity of $\phi$ on cut $d$. To obey \eqref{eq:ampere}, the difference between $\phi$ on each side of the discontinuity simply needs to be defined as the enclosed current $I_{\text{enc}}$ inside loop $C$. 
	
		\begin{figure*}[tb]
		\centering
		\includegraphics[width=\linewidth]{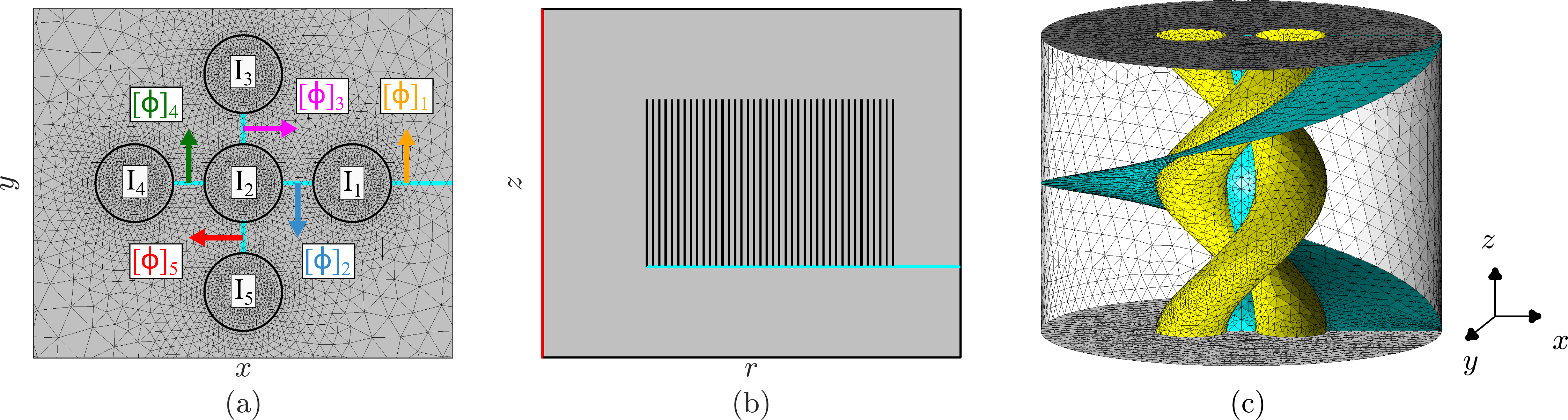}
		\caption{Geometries studied in this work. (a) Simple \twoD case showing five infinitely long superconducting wires carrying current amplitudes of $I_1=60~$A, $I_2=-80~$A, $I_3=40~$A, $I_4=20~$A and $I_5=80~$A. The direction of discontinuities [$\phi$]$_i$ are shown by the arrows and their amplitudes are detailed in the main text. (b) \twoD axisymmetric geometry of a superconducting pancake coil. The red line represents the symmetry axis. The tapes are enlarged for visualization purposes. (c) \threeD geometry of superconducting twisted filaments. In all figures, the cyan lines (surfaces) in \twoD (\threeD) represent the cuts used to impose the discontinuity in~$\phi$. The mesh used for the FEM simulations is shown for all geometries except for the pancake coils, where the mesh elements are too small to be clearly visualized.}
		\label{fig:geom}
	\end{figure*}
	
	Two steps are needed to implement the thin cut in COMSOL. First, a regular line segment (or plane in 3-D) must be defined in the geometry where the cut will be imposed. Then, the scalar potential discontinuity is applied on the line segment with the use of a \textit{Magnetic Scalar Potential Discontinuity} node in the predefined \textit{Magnetic Fields No Currents} (MFNC) module. This simply duplicates the number of DOF on the line (or surface) and constrains $\phi$ on each side of the cut to be equal to the specified current. A sample model will be uploaded to the htsmodelling.com website for further details \cite{htsmodeling}.
	
	Careful attention must be paid to the sign of the applied discontinuity in \eqref{eq:discontinuity}, since the resulting current's sign depends on the direction of contour $C$ and the location of $d^+$ and $d^-$ with respect to the cut. The direction of contour $C$ can easily be established by using the right-hand rule on the coordinate system of the model. Moreover, the direction from $d^{+}$ to $d^{-}$ is graphically shown by an arrow through the edge or surface in COMSOL. If the direction of contour $C$ is in the same direction as $d^+$ to $d^-$ through the cut, a positive (negative) discontinuity will result in a positive (negative) enclosed current. On the other hand, if the direction of contour $C$ is in the opposite direction of $d^+$ to $d^-$, then a positive (negative) jump will lead to a negative (positive) enclosed current.

	Note that since the magnetic field is obtained by taking the negative gradient of $\phi$, the discontinuity of $\phi$ has no influence on the magnetic field results. In addition, there exist an infinite number of valid cuts, but the length of the cut should be minimized because the DOFs are duplicated along each node of the cut in order to impose the discontinuity in $\phi$.

	Finally, although using the magnetic scalar potential as opposed to the full vector field greatly reduces the number of DOFs in non-conducting domains, the need for imposing cuts in complex topologies can be very tricky. Solutions proposed include algorithms for automatically generating the proper cuts and using thick cuts \cite{ren2002,leonard1993,anhtuanphung2005,dlotko2015,dlotko2009,stockrahm2019,stenvall2014,lahtinen2015}. However, these methods are not currently implementable in COMSOL Multiphysics. We therefore restrict ourselves to relatively simple geometries in this paper, since the cuts must be manually added to the geometries.

\section{Application Examples}

In order to evaluate the potential of the \mbox{\hphi-formulation} with thin cuts, we model three different cases involving superconducting filaments and tapes. We begin with a simple demonstration of five infinitely long superconducting filaments carrying different transport currents in \twoD. In the second case, we model a pancake coil carrying a transport current in a \twoD axisymmetric simulation. Finally, we model twisted superconducting cables carrying transport currents. The three geometries studied are shown in Fig.~\ref{fig:geom}. Note that the cuts made on each geometry were chosen to minimize their lengths, but other cuts could be used and would lead to the same results. The line segments (planes in 3-D) representing the cuts are also added to the geometries in all formulations considered for consistency, but they are only used for the physics in the \hphi-formulation.

\subsection{\twoD superconducting filaments}

To demonstrate how the cuts are implemented, we begin by modeling a simple geometry consisting of five infinitely long superconducting filaments carrying different transport currents. The filaments are separated by a distance of 1~mm and have a cross-sectional area of 0.1~mm$^2$, so that their critical current is 100~A. The geometry considered is shown in Fig.~\ref{fig:geom}(a), where the filaments carry arbitrarily chosen current amplitudes of $I_1=60~$A, $I_2=-80~$A, $I_3=40~$A, $I_4=20~$A and $I_5=80~$A, with a frequency of $f=50$~Hz. The arrows on each cut in Fig.~\ref{fig:geom}(a) also appear graphically in COMSOL Multiphysics and demonstrate the direction of $d^+$ and $d^-$ along the cut.

The applied discontinuities [$\phi$]$_3$, [$\phi$]$_4$ and [$\phi$]$_5$ are simply given by the applied currents in each respective filament because the filaments are only connected to one cut. However, filaments with currents $I_1$ and $I_2$ are connected to multiple cuts so that the discontinuities must be carefully determined. Taking a loop around the filament with current $I_1$, both [$\phi$]$_1$ and [$\phi$]$_2$ contribute to the discontinuity, so that their sum must be equal to $I_1$. Moreover, considering a closed curve around the center filament, the contributions from [$\phi$]$_2$, [$\phi$]$_3$, [$\phi$]$_4$ and [$\phi$]$_5$ are negative since $C$ is counter-clockwise according to the right-hand rule and the corresponding arrows are in the opposite direction of $C$. The potential discontinuities are related to the currents in the filaments through the following matrix:
\[
\begin{bmatrix}
1 & 1 & 0 & 0 & 0\\
0 & -1 & -1 & -1 & -1\\
0 & 0 & 1 & 0 & 0\\
0 & 0 & 0 & 1 & 0\\
0 & 0 & 0 & 0 & 1
\end{bmatrix}
\begin{bmatrix}
[\phi]_1\\
[\phi]_2\\
[\phi]_3\\
[\phi]_4\\
[\phi]_5
\end{bmatrix}
=
\begin{bmatrix}
I_1\\
I_2\\
I_3\\
I_4\\
I_5
\end{bmatrix}
\sin(2\pi f t)
\]

The same procedure can be adopted for any current amplitude and/or waveform. The potential discontinuities in each cut are simply a linear combination of the currents imposed in the conductors.

\begin{figure}[tb]
	\centering
	\includegraphics[width=\linewidth]{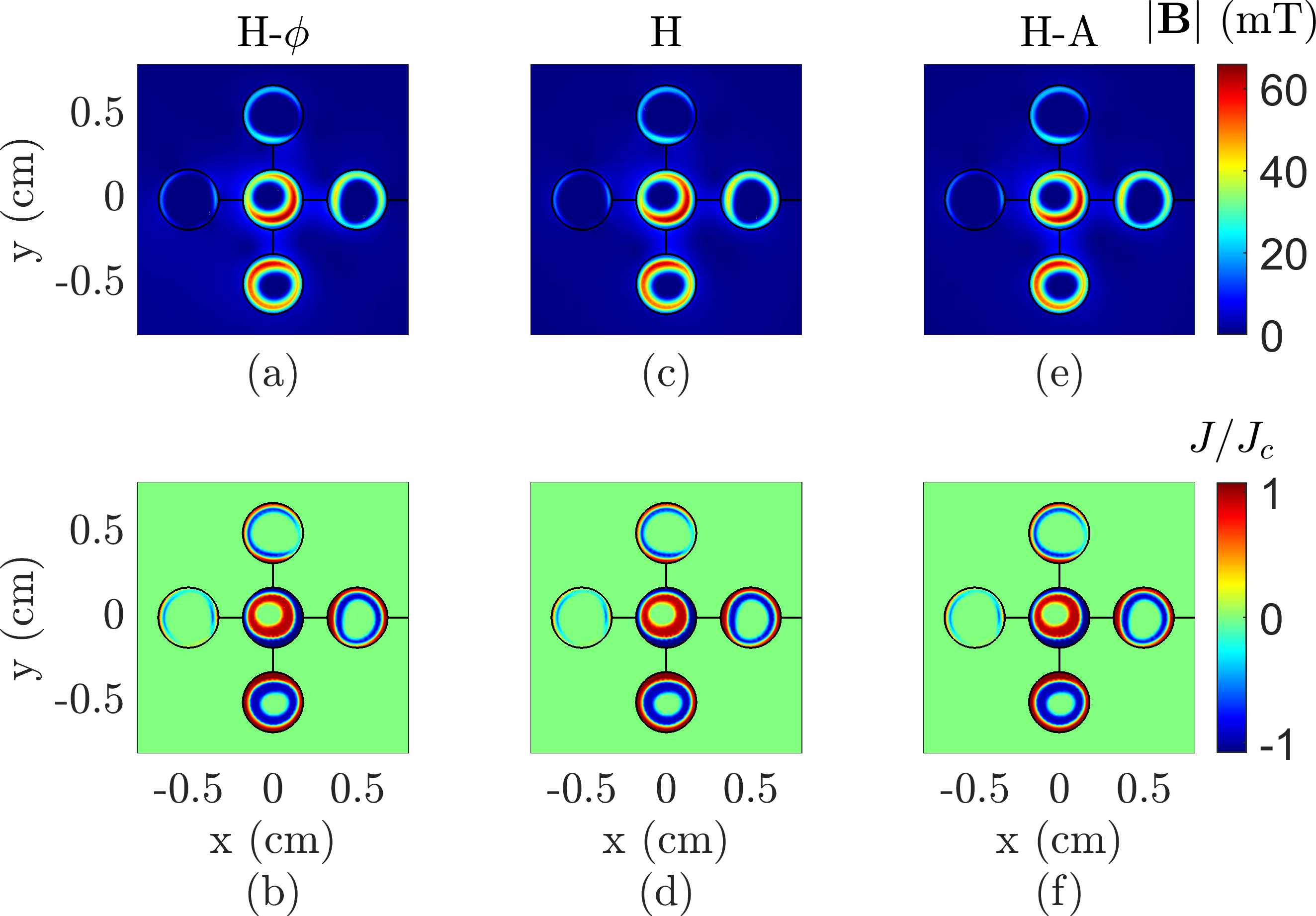}
	\caption{Simulation results for the infinitely long \twoD superconducting filaments simulated with the \hphi and H-formulations. The top row shows the magnetic flux density simulated with the \hphi (a), H (c), and H-A (e) formulations. The bottom row shows the current density normalized by $J_c$ simulated with the \hphi (b), H (d), and H-A (f) formulations.}
	\label{fig:2D}
\end{figure}

The resulting magnetic flux density and normalized current density after one cycle are shown in Figs.~\ref{fig:2D}(a) and (b), respectively. We simulate the filaments with 9,678 linear elements in the whole simulation space, corresponding to 9,418 DOFs using the \mbox{\hphi-formulation}.

To assess the accuracy of the results, we compare them with the standard H-formulation using the same number of mesh elements, corresponding to a total of 14,561 DOFs. In this case, the air domains are simulated with \eqref{eq:H}, with a high resistivity of 1~$\Omega$m, such that negligible currents are induced in the air. The transport currents are applied with algebraic constraints:
\begin{equation}
\int_{\Omega_i}\nabla\times\mathbf{H}~\textrm{d}\Omega_i=\mathbf{I}_i,
\label{eq:constraint}
\end{equation}
where $\Omega_i$ and $\mathbf{I}_i$ are the cross-section and transport current of filament $i$, respectively. The magnetic flux density and normalized current density distributions calculated with the H-formulation are shown in Fig.~\ref{fig:2D}(c) and (d), respectively. We find good agreement between the two formulations.

In order to evaluate the computational advantage of using thin cuts vs. algebraic constraints to impose a transport current, we also compare the \mbox{\hphi-formulation} with the H-A-formulation \cite{brambilla2018}. The H-A-formulation uses the magnetic vector potential in the air domains, such that the current density is zero and the dependent variable is a scalar (in \twoD). Therefore, the main difference between the two formulations is the coupling between the mixed formulations and the method used to impose the current inside the conducting domains. The \mbox{H-A-formulation} uses the algebraic constraint in \eqref{eq:constraint}, whereas the \mbox{\hphi-formulation} imposes a discontinuity in $\phi$. 

When using the same number of linear elements in the \mbox{H-A-formulation}, the number of DOFs is 9,379, which is 39 fewer DOFs than that of the \mbox{\hphi-formulation} since cuts are not used. The magnetic flux density and current density distributions calculated with the \mbox{H-A-formulation} after one cycle are shown in Fig.~\ref{fig:2D}(e) and (f), respectively. The \hphi, H-A, and H formulations are all found to be visually equivalent.

\begin{figure}[tb]
	\centering
	\includegraphics[width=\linewidth]{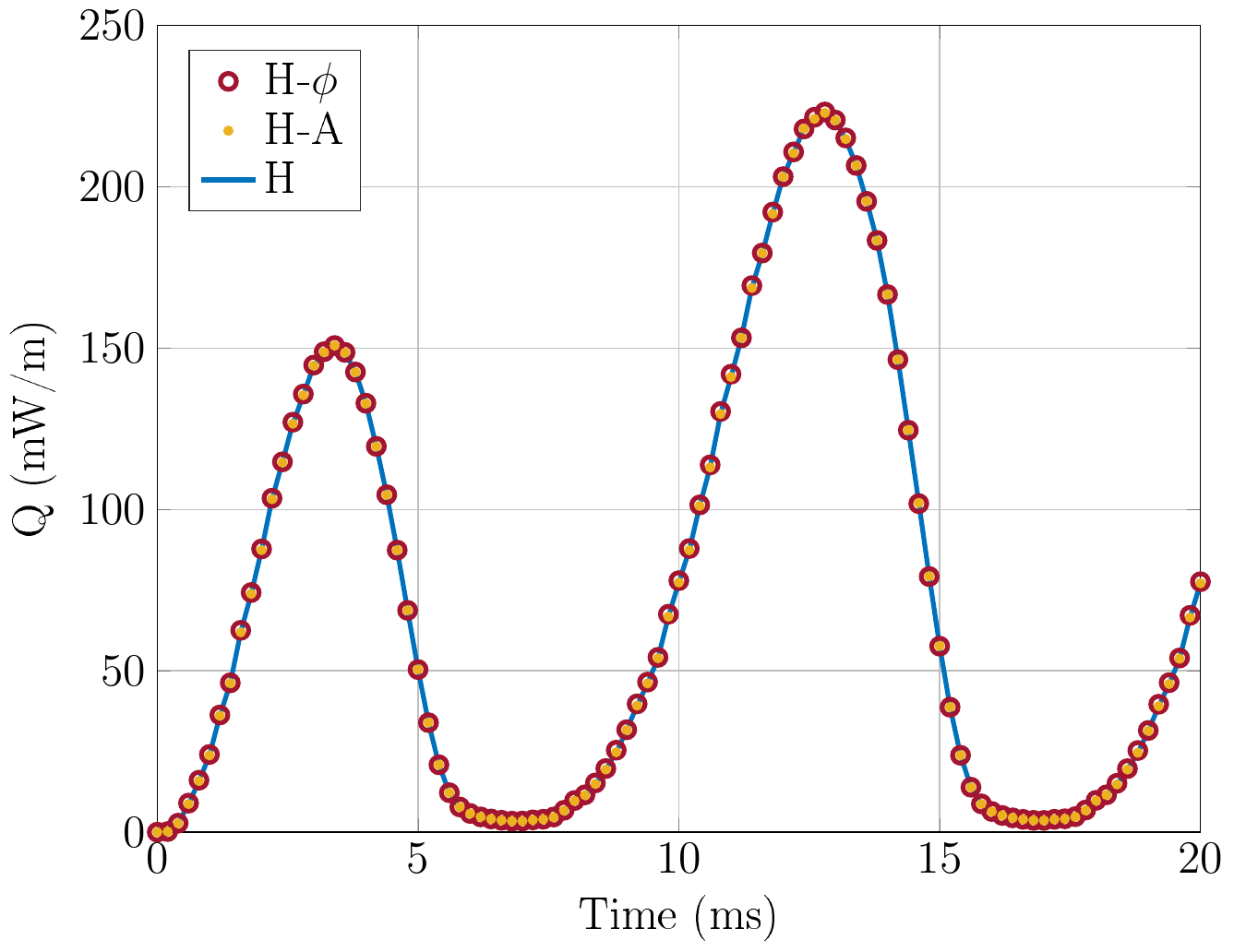}
	\caption{AC losses in all five filaments of Fig.~\ref{fig:geom}(a) computed with the \hphi, H-A and H formulations.}
	\label{fig:2D_ac}
\end{figure}

Fig.~\ref{fig:2D_ac} shows a comparison of the instantaneous AC losses in all the filaments combined, calculated with all three formulations, using:
\begin{equation}
Q=\int_{\Omega}\mathbf{E}\cdot\mathbf{J}~\textrm{d}\Omega,
\end{equation}
where $\Omega$ is the superconducting domain. We find excellent agreement, with the difference between formulations remaining below 850~$\mu$W/m at every time step.

As a final comparison, we evaluate the variance of the current density distribution between formulations by calculating the coefficient of determination \cite{berrospe-juarez2021}. This is given by:
\begin{equation}
R^2=1-\frac{\int_0^{\tau}\int_{\Omega_{SC}}\left(\mathbf{J_{H}}-\mathbf{J_{H-x}}\right)^2\textrm{d}\Omega_{SC}\textrm{d}t}{\int_0^{\tau}\int_{\Omega_{SC}}\left(\mathbf{J_{H}}-\overline{\mathbf{J_{H}}}\right)^2\textrm{d}\Omega_{SC}\textrm{d}t},
\label{eq:R2}
\end{equation}
where $\tau$ is the time for one cycle, $\Omega_{SC}$ is the superconducting domain, $\mathbf{J_H}$ is the current density calculated with the H-formulation, $\mathbf{J_{H-x}}$ is the current density calculated with the \hphi or the \mbox{H-A-formulation}, and $\overline{\mathbf{J_{H}}}$ is the current density obtained with the H-formulation averaged over $\Omega_{SC}$. This coefficient enables us to quantitatively evaluate the difference of the current density between the formulations over the whole superconducting domain and over all time steps. Note that $R^2=1$ implies that $\mathbf{J_H}$ and $\mathbf{J_{H-x}}$ are equivalent. Calculated over all five filaments, we obtain $R^2=0.9969$ for the \mbox{\hphi-formulation} and $R^2=0.9999$ for the \mbox{H-A-formulation}, demonstrating that all formulations are nearly identical.

The computation times needed to simulate one cycle are 4.05~minutes, 5.18~minutes, and 11.53~minutes for the \hphi, H-A, and H-formulations, respectively. Thus, the H-A and \hphi formulations are clearly advantageous in this simple \twoD case. However, we find that the constraints used to impose the current in the H-A and H-formulation are more computationally expensive than the cuts needed in the \mbox{\hphi-formulation}. Indeed, even with 39 fewer DOFs, the \mbox{H-A-formulation} is still 22\% slower than the \mbox{\hphi-formulation}. See Table~\ref{tbl:summary} for a summary of the results.

\subsection{Pancake coil}

We model a superconducting pancake coil by simulating its cross-section in a \twoD axisymmetric geometry, shown in Fig.~\ref{fig:geom}(b). The studied coil consists of 40 turns of superconducting tapes with a thickness of 1~$\mu$m and a width of 4~mm, separated by a distance of 150~$\mu$m. The coil has an inner radius of 3~cm with a transport current of $I_0=96$~A at $f=50$~Hz, corresponding to 80\% of the critical current. As done in \cite{zhang2012}, we only consider the superconducting layer in the tapes; other layers such as the substrate and copper layers are neglected for simplicity. We use two mesh elements along the thickness of each tape and 50 mesh elements along their width, corresponding to a total of 38,060 linear elements in all domains.

As shown in Fig.~\ref{fig:geom}(b), we chose the cuts to be between each superconducting tape in order to minimize their length. Consequently, the discontinuities applied on the cuts must be the sum of all transport currents encompassed by the cut. For example, the right outermost tape has the only cut ($d1$) connected to the air boundary, such that the cut must have a discontinuity of $[\phi]_{d1}=40I_0\sin(2\pi f t)$. The neighboring cut ($d2$) must then have a discontinuity of $[\phi]_{d2}=39I_0\sin(2\pi f t)$ and so on, until the current in each tape contributes to the total $\phi$ discontinuity in the associated cuts.

\begin{figure}[tb]
	\centering
	\includegraphics[width=\linewidth]{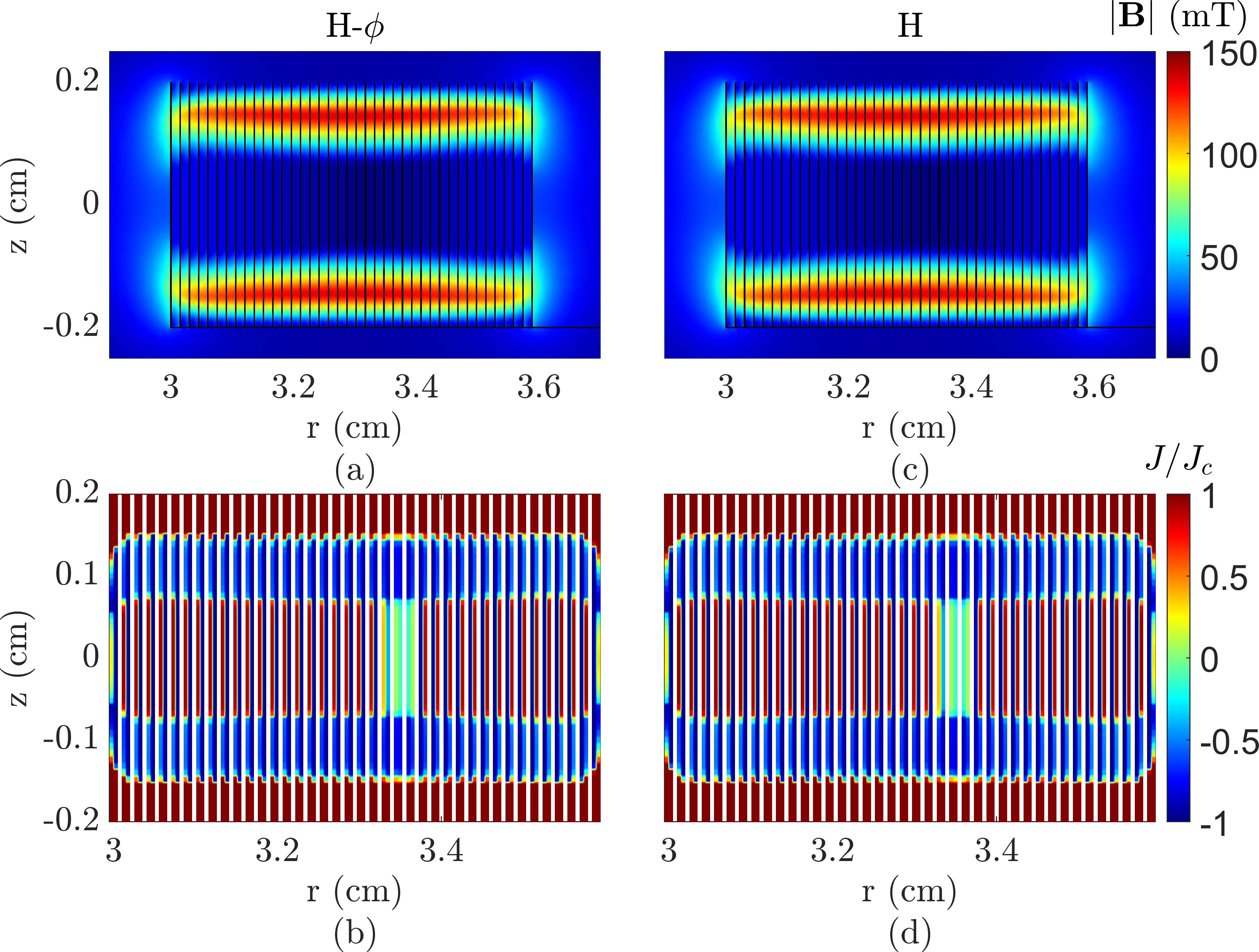}
	\caption{Simulation results for the \twoD axisymmetric superconducting pancake coil simulated with the \hphi and H-formulations. (a) Norm of the magnetic flux density and (b) normalized current density calculated with the \mbox{\hphi-formulation} at the end of the first cycle. (c) Norm of the magnetic flux density and (d) normalized current density calculated using the \mbox{H-formulation}. The tapes in (b) and (d) are scaled in the $r$-direction by a factor of 100 to better visualize the current density distribution.}
	\label{fig:pancake}
\end{figure}

\begin{figure}[tb]
	\centering
	\includegraphics[width=\linewidth]{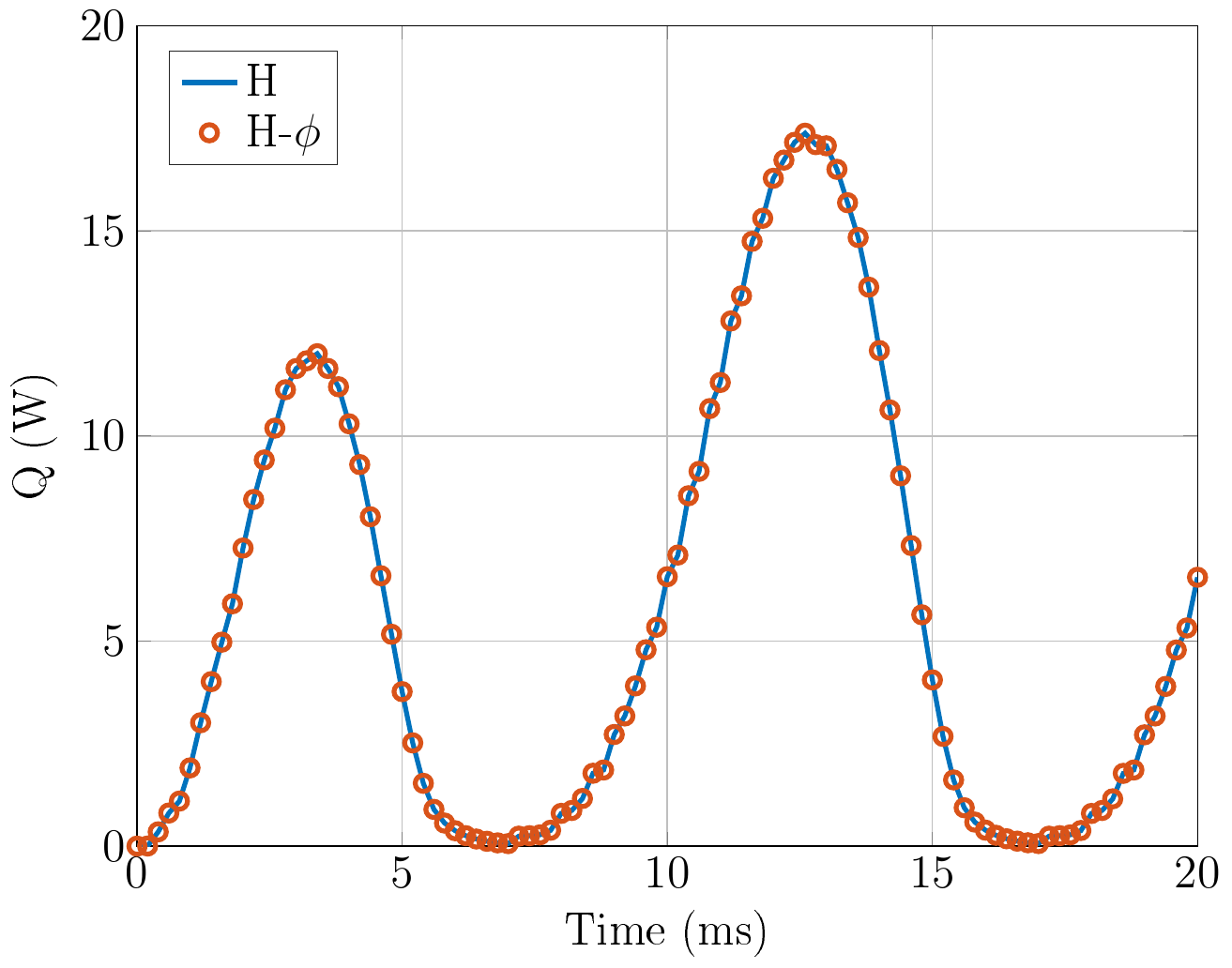}
	\caption{AC losses in the pancake coil computed with the \hphi and H formulations.}
	\label{fig:pancake_ac}
\end{figure}

The resulting magnetic flux density and normalized current densities at the end of the first cycle are shown in Fig.~\ref{fig:pancake}(a) and (b), respectively. When compared with the \mbox{H-formulation} simulations of Fig.~\ref{fig:pancake}(c) and (d), we find remarkable agreement between formulations. The AC loss computations of Fig.~\ref{fig:pancake_ac} also show that global quantities agree, with the difference between formulations remaining below 5~$\mu$W at every time step.

In order to quantitatively compare the current density distributions obtained with both formulations, we calculate the $R^2$ coefficient of \eqref{eq:R2}. In this case, we take $\Omega_{SC}$ as the cross-section of all the tapes, so that the coefficient of determination is calculated only on the \twoD axisymmetric domain. We obtain $R^2=1.0000$, meaning that the result of the two formulations are exactly the same up to four decimal places. 

\begin{figure*}
	\centering
	\includegraphics[width=\linewidth]{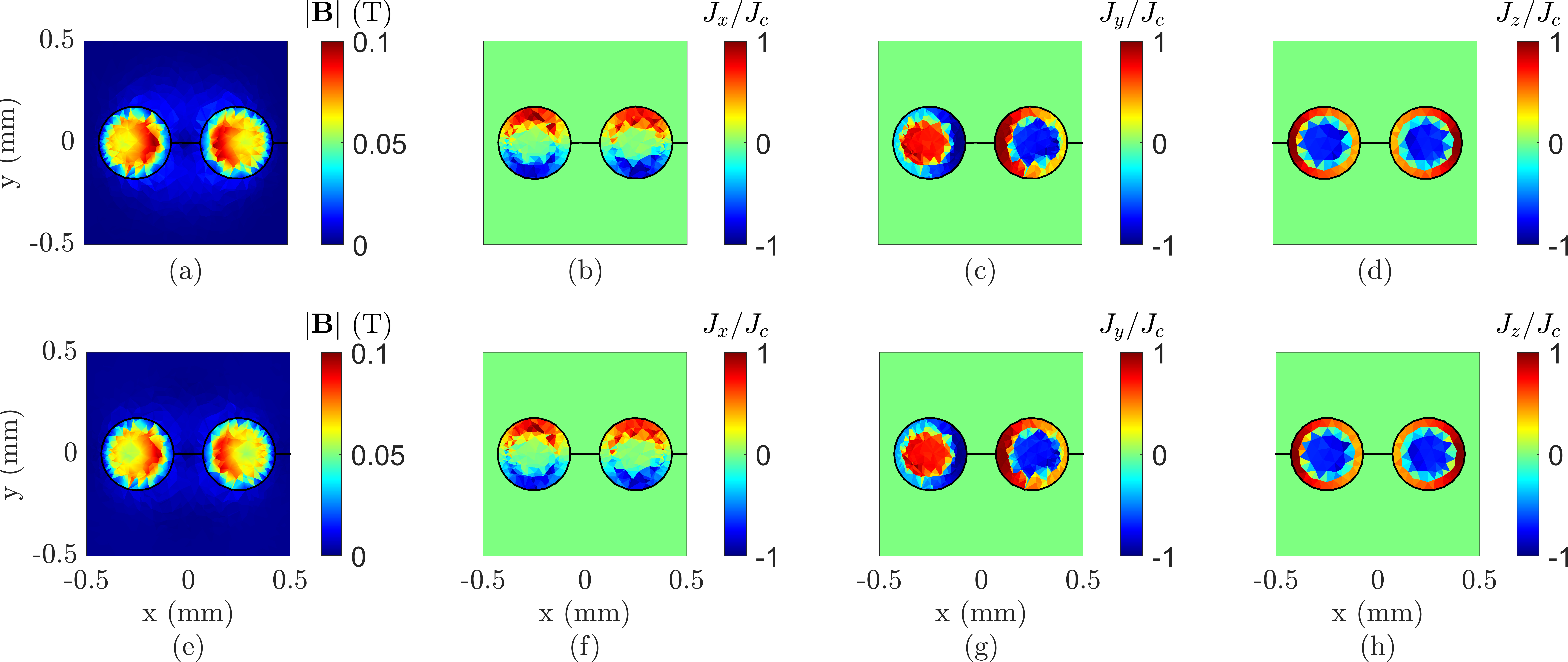}
	\caption{Simulation results for the twisted superconducting filaments in the $x$-$y$ plane for $z=0.75$~mm (center of the geometry). Top (bottom) row shows results from the \hphi (H) formulation. (a) and (e) Norm of the magnetic flux density. (b) and (f) Current density in the $x$-direction. (c) and (g) Current density in the $y$-direction. (d) and (h) Current density in the $z$-direction.}
	\label{fig:twisted}
\end{figure*}

The advantage of using the \mbox{\hphi-formulation} with cuts is notable when comparing the computation times in the pancake coil model. In this case, we use 38,060 linear elements in both formulations, corresponding to 29,383 DOFs and 59,140 DOFs for the \hphi and \mbox{H-formulation}, respectively. Since there are many elements in the HTS domains, we found that the constraint \eqref{eq:constraint} leads to extremely long computation times. We therefore use:
\begin{equation}
\int_{\partial\Omega_{i}}\mathbf{H}\cdot\text{d}\mathbf{l}=I
\label{eq:constraint_boundary}
\end{equation}
to constrain the current in the H-formulation of the pancake coils. Consequently, the computation times for the \hphi and \mbox{H-formulation} are 35~minutes and 165~minutes, respectively. When using \eqref{eq:constraint}, the computation time is up to 1,195~minutes with the H-formulation. These time differences show that the constraints \eqref{eq:constraint} and \eqref{eq:constraint_boundary} used in the \mbox{H-formulation} are computationally expensive when compared to the use of cuts. Indeed, \eqref{eq:constraint} and \eqref{eq:constraint_boundary} must be obeyed for all elements in the tapes or at the boundaries of the tapes, while the constraint imposed in the \mbox{\hphi-formulation} must only be obeyed on the nodes of the cuts. As we have previously shown, without any imposed current, the \mbox{\hphi-formulation} is approximately twice as fast as the \mbox{H-formulation} in \twoD \cite{arsenault2021}. Here, we find that the \mbox{\hphi-formulation} is nearly 5 times more efficient when a net current is imposed. See Table.~\ref{tbl:summary} for a summary of the results.

Note that we do not compare with the \mbox{H-A-formulation} in this model since we found that spurious oscillations in the current density appear due to the coupling between H and A, similar to those observed in the \mbox{T-A-formulation} \cite{berrospe-juarez2019}. The H-A formulation requires enriched elements in either the H or A physics to avoid these oscillations \cite{dular2021}, which would necessarily be slower than the \hphi model. In addition, the \mbox{H-A-formulation} requires the same current constraints as the \mbox{H-formulation}, leading to longer computation times than the \mbox{\hphi-formulation}.

\subsection{Twisted filaments}

To demonstrate a more complex geometry modeled using thin cuts in \threeD, we model a set of twisted superconducting filaments, as shown in Fig.~\ref{fig:geom}(c). The filaments have a cross-sectional area of 0.1~mm$^2$ with a critical current of 100~A. They are separated by a center-to-center distance of 0.5~mm and have a twist pitch of 4~mm. We impose a sinusoidal transport current of $I_0=80$~A at $f=50$~Hz in each filament. 

In order to reduce the span of the cuts, we take one cut between both filaments and one cut that reaches the external air boundary. As such, the central cut ($c$) must contain a discontinuity in $\phi$ of $[\phi]_{c}=I_0\sin(2\pi f t)$, while the external cut ($e$) must have a discontinuity of $[\phi]_{e}=2I_0\sin(2\pi f t)$.

We add a periodic boundary condition (PBC) at each end of the domain to obtain the correct field behavior. However, careful attention must be paid to the PBC, since it makes the air domain multiply connected \cite{bettini2017}. Indeed, a curve going straight from the bottom face to the top face of the air domain will acquire a discontinuity in $\phi$ equal to $[\phi]_{e}$, even though it does not surround a conductor. Therefore, the PBC on $\phi$ must read:
\begin{equation}
\phi_b-\phi_t=[\phi]_e,
\end{equation}
where $\phi_b$ and $\phi_t$ represent the magnetic scalar potential on the bottom and top faces of the air domain, respectively. This modified PBC can be thought of as adding another horizontal thin cut to make the air domains simply connected again. In addition, due to the nature of how the PBC are implemented in COMSOL, the discontinuity in $\phi$ is not properly projected from one face to another. We solve this issue by using a weak form PBC, meaning that the constraint is applied with the use of Lagrange multipliers in the weak formulation.

The magnetic flux density and current density distributions obtained with both the \hphi and \mbox{H-formulation} are shown in Fig.~\ref{fig:twisted} at the end of one current cycle. In both cases, we use 283,493 linear elements, corresponding to 96,882 DOFs and 337,639 DOFs in the \hphi and \mbox{H-formulation}, respectively.

\begin{figure}[tb]
	\centering
	\includegraphics[width=\linewidth]{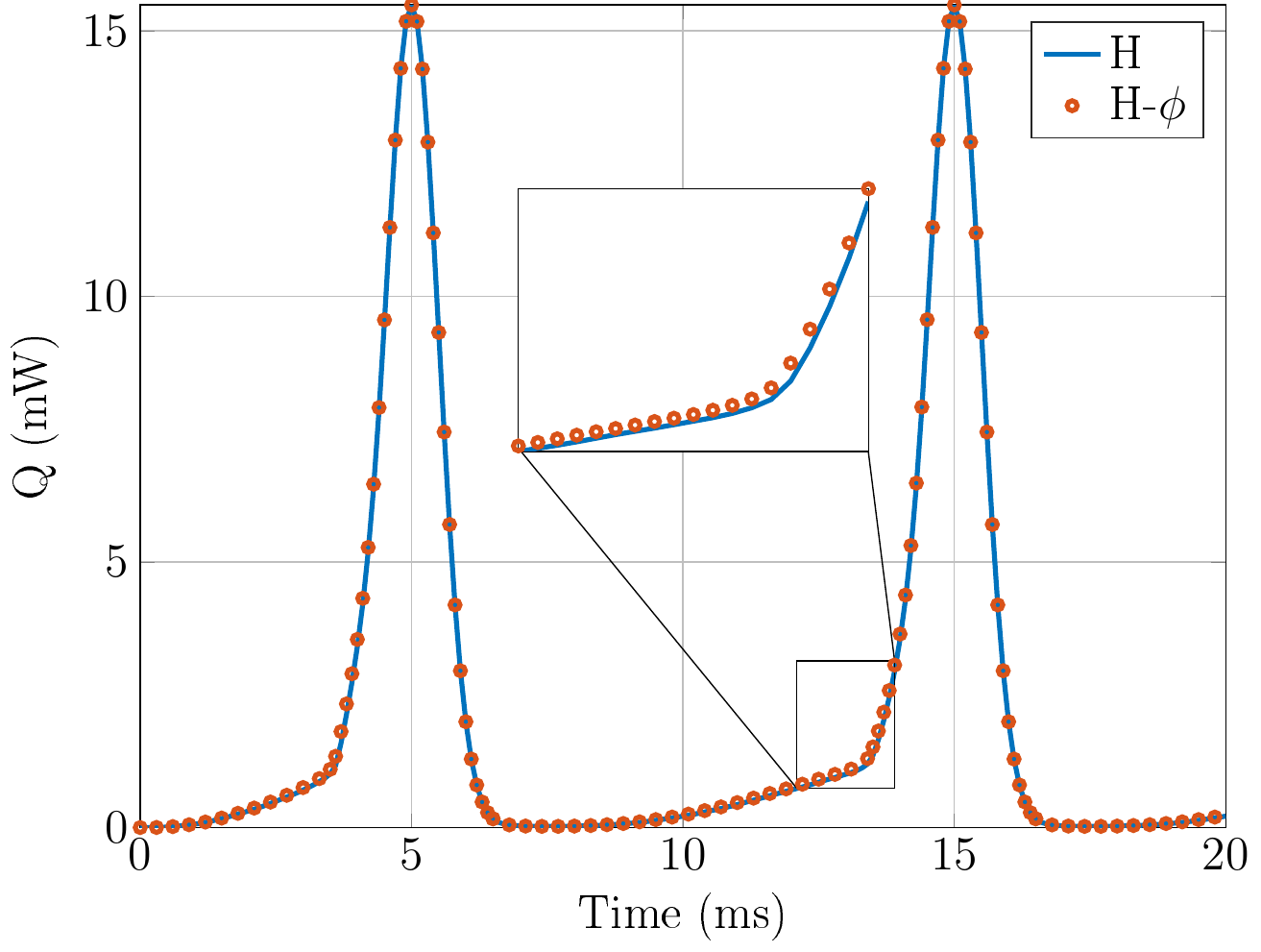}
	\caption{AC losses in both twisted superconducting filaments computed with the H and \hphi formulations. The inset shows a closeup of the plot in the region of largest variation between the two formulations.}
	\label{fig:twisted_ac}
\end{figure}

\begin{table*}[tb]
	\centering
	\caption{Summary of the comparison between formulations for the different applications studied in this work}
	\setlength{\tabcolsep}{3pt}
	\begin{tabular}{ l  l  l  l l l}
		\toprule
		Application & Formulation & Elements & DOFs &  Time (min) & $R^2$ \\
		\midrule
		
		\multirow{3}{*}{\twoD filaments} & \hphi & 9,678 & 9,418 & 4.05 & 0.9969\\
		{}	& H-A & 9,678 & 9,379 & 5.18 & 0.9999\\
		{}	& H & 9,678 & 14,561 & 11.53 & ---\\\\
		
		\multirow{2}{*}{Pancake coil} & \hphi & 38,060 & 29,383 & 35 & 1.0000\\
		{}	& H & 38,060 & 59,140 & 165 & ---\\\\
		
		\multirow{2}{*}{Twisted filaments} & \hphi & 283,493 & 96,882 & 48 & 0.9934\\
		{}	& H & 283,493 & 337,639 & 356 & ---\\\\
		
		\bottomrule		
	\end{tabular}
	\label{tbl:summary}
\end{table*}

Although the current densities are visually similar between the two formulations, as shown in Fig.~\ref{fig:twisted}(b)-(d) and (f)-(h), the magnetic flux density of Fig.~\ref{fig:twisted}(a) and (e) is slightly higher ($\sim$6\%) in the \mbox{\hphi-formulation}. Several reasons might explain this discrepancy. First, the current can leak into the air domains when using the \mbox{H-formulation} due to the dummy resistivity \cite{stenvall2014}. We calculated the total current inside both filaments at different $z$ values and found that the current simulated in the \mbox{H-formulation} drops by $3\times10^{-7}$\% when compared to the imposed current in \eqref{eq:constraint}. Therefore, the leaked current is not the main factor affecting the difference in field between the two formulations. A more reasonable explanation is that the divergence-free condition of the field is not explicitly imposed in the \mbox{H-formulation}, which can lead to an accumulation of errors at every time step \cite{zermeno2013a}. In addition, the coupling between the H and $\phi$ physics can also lead to errors \cite{arsenault2021}. 

Nevertheless, the AC losses computed with both formulations, shown in Fig.~\ref{fig:twisted_ac}, are in very good agreement. We find that the difference between formulations remains below 217~$\mu$W for all time steps. Fig.~\ref{fig:twisted_ac} also shows a closeup of the plot where the largest variation between formulations is present. Apart from this region and near the 3~ms time stamp, the AC losses calculated in both formulations are visually indistinguishable. In addition, we calculate a coefficient of determination of $R^2=0.9934$, meaning that the current density distributions are in excellent agreement. 

For this model, the \mbox{\hphi-formulation} takes 48~minutes to compute, while the \mbox{H-formulation} takes 356~minutes, showing a seven-fold increase in computation time. See Table.~\ref{tbl:summary} for a summary of the results.

Note that we do not compare with the \mbox{H-A-formulation} in this case since A is a vector in \threeD and would therefore not be more advantageous than the \mbox{\hphi-formulation}.

\section{Conclusion}

In this work, we demonstrated that thin cuts are a simple and efficient method of making non-conducting domains simply connected in the \mbox{\hphi-formulation} to model superconductors in COMSOL Multiphysics. These cuts can be implemented in commercial FEM software such as COMSOL Multiphysics, since they do not require additional basis functions or low-level programming. 

We simulated three different models employing HTS filaments and tapes. First, we demonstrated how thin cuts are used in a simple \twoD case of five infinitely long superconducting filaments carrying different transport currents. In this case, the H, \hphi and H-A formulations were compared, since the \mbox{H-A-formulation} is also beneficial in \twoD. When compared with the \mbox{H-formulation}, the R$^2$ coefficient is 0.9969 and 0.9999 for the \hphi and \mbox{H-A-formulation}, respectively. Therefore, the current density distributions were found to be nearly identical, with the computation time of the \mbox{\hphi-formulation} being nearly three times faster than the \mbox{H-formulation}. We found that the \mbox{H-A-formulation} is 22\% slower than the \mbox{\hphi-formulation}, even though it uses 39 fewer DOFs. Thus, we concluded that the use of thin cuts in the \mbox{\hphi-formulation} is more efficient than the constraint used in the H and H-A formulations to impose the current.

We then modeled a superconducting pancake coil in a \twoD axisymmetric simulation. In this model, the H and \hphi formulations were found to be identical, since the R$^2$ value was 1.0000. However, the time taken to simulate one cycle was nearly five times faster when using the \mbox{\hphi-formulation}.

Finally, we considered a \threeD model consisting of a set of twisted superconducting filaments carrying a transport current. The $R^2$ value was found to be 0.9934 between formulations, yet the \mbox{\hphi-formulation} was more than seven times faster than the H formulation.

Ultimately, although the addition of thin cuts makes the \mbox{\hphi-formulation} slightly more challenging to implement than the simple \mbox{H-formulation}, it provides a drastic improvement in terms of computation times with respect to the H and H-A-formulations for an equivalent accuracy. One limitation of this method when implemented in COMSOL Multiphysics is that the cuts must be implemented manually, such that only relatively simple geometries can be modeled. Fortunately, this is not a limit for most practical cases.

	\section{Acknowledgements}
	
	This work was supported in part by the Fonds de recherche du Qu\'ebec --- Nature et Technologies (FRQNT), TransMedTech Institute and its main funding partner, the Canada First Research Excellence Fund and Coordena\c{c}\~ao de Aperfei\c{c}oamento de Pessoal de N\'{i}vel Superior – Brazil (CAPES) - Finance code 001. 
	

	
	
	%
\begingroup
\raggedright
\bibliography{Bibliography}
\bibliographystyle{IEEEtran}
\endgroup
	
	%
	
	\begin{IEEEbiographynophoto}{Alexandre Arsenault}
		received a B.Sc. in physics from McGill University, Montr\'eal, QC, Canada, in 2016. He also received a M.Sc. in physics from McMaster University, Hamilton, ON, Canada, in 2018. He is currently pursuing a Ph.D. degree in biomedical engineering at Polytechnique Montr\'eal under the supervision of Prof. Fr\'ed\'eric Sirois. His research interests include the characterization and simulation of high-temperature superconductors and magnetic drug delivery.
	\end{IEEEbiographynophoto}

\begin{IEEEbiographynophoto}{Bruno de Sousa Alves}
	received the B.S. in electrical and industrial engineering and the M.Sc. degree in electrical engineering from the Federal University of Santa Catarina, Florianópolis, Brazil, in 2014 and 2016, respectively. He is currently pursuing a Ph.D. degree in electrical engineering at Polytechnique Montr\'eal under the supervision of Prof. Fr\'ed\'eric Sirois and Prof. Marc Laforest. His research interests include nonlinear modeling and analysis of electromagnetic devices, with emphasis on modeling conducting and ferromagnetic thin layers in harmonic and time-transient regimes.
\end{IEEEbiographynophoto}

\begin{IEEEbiographynophoto}{Fr\'ed\'eric Sirois}(S'96--M'05--SM'07)
	received the B.Eng. degree in electrical engineering from Universit\'e de Sherbrooke, Sherbrooke, QC, Canada, in 1997, and the Ph.D. degree in electrical engineering from Polytechnique Montr\'eal, Montr\'eal, QC, Canada, in 2003.
	From 1998 to 2002, he was affiliated as a Ph.D. scholar with the Hydro-Qu\'ebec's Research Institute (IREQ), where he was a Research Engineer from 2003 to 2005. In 2005, he joined Polytechnique Montr\'eal, where he is currently Full Professor. His main research interests are i) the characterization and modeling of electric and magnetic properties of materials, ii) modeling and design of electromagnetic and superconducting devices, and iii) integration studies of superconducting equipment in power systems. He is a regular reviewer for several international journals and conferences.
\end{IEEEbiographynophoto}

	
	

\end{document}